\documentclass[amsmath,superscriptaddress,preprint,aps]{revtex4-2}
\usepackage{graphicx}% Include figure files
\usepackage{dcolumn}% Align table columns on decimal point
\usepackage{bm}% bold math
\usepackage{siunitx}
\usepackage{color, soul}
\setstcolor{red}
\usepackage[mathlines]{lineno}% Enable numbering of text and display math
\begin{document}
	\title{Exceptional point proximity-driven mode-locking in coupled microresonators}

	\author{Riku Imamura}
 	\affiliation{Department of Electronics and Electrical Engineering, Faculty of Science and Technology, Keio University, Yokohama, 223-8522, Japan}
	\author{Shun Fujii}
 	\affiliation{Department of Physics, Faculty of Science and Technology, Keio University, Yokohama, 223-8522, Japan}
  	\author{Ayata Nakashima}
 	\affiliation{Department of Electronics and Electrical Engineering, Faculty of Science and Technology, Keio University, Yokohama, 223-8522, Japan}
	\author{Takasumi Tanabe}
	\email[Corresponding author. ]{takasumi@elec.keio.ac.jp}
 	\affiliation{Department of Electronics and Electrical Engineering, Faculty of Science and Technology, Keio University, Yokohama, 223-8522, Japan}
  \date{\today}% It is always \today, today,
	% but any date may be explicitly specified
\begin{abstract}
We show theoretically and numerically that mode-locking is feasible with a coupled-cavity system with gain and loss, notably, without any natural saturable absorber. We highlight that in the vicinity of the exceptional point, system $Q$ exhibits substantial modulation even with minor refractive index changes and a minimal Kerr effect contribution. Leveraging this unique behaviour, we propose an unprecedented approach wherein the lossy auxiliary cavity functions as an efficient artificial saturable absorber, thus facilitating mode-locking. This approach is not only novel, but also presents considerable advantages over conventional systems where both gain and saturable absorption are contained within a single microcavity. These benefits include reduced operational power and ease of post-adjustment, achievable through the manipulation of the coupling strength between the two microcavities.
\end{abstract}
	\maketitle
 %%%%%%%%%%%%%%%%%%%%%%%%%%%%%%%%%%%%%%
 %%%%%%%%%%%%%%%%%%%%%%%%%%%%%%%%%%%% To HERE

 %  Optica template (comment out if unnecessary)
%%%%%%%%%%%%%%%%%%%%%%%%%%%%%%%%%%%%%%%%%

% \documentclass{optica-article}
% \journal{opticajournal} % for journals or Optica Open
% \articletype{Research Article}
% \usepackage{lineno}
% \linenumbers % Turn off line numbering for Optica Open preprint submissions.
% \begin{document}

% \title{Exceptional point proximity-driven mode-locking in coupled microresonators}

% \author{Riku Imamura,\authormark{1} Shun Fujii,\authormark{2,} Ayata Nakashima,\authormark{1}}

% \address{\authormark{1}Peer Review, Publications Department, Optica Publishing Group, 2010 Massachusetts Avenue NW, Washington, DC 20036, USA\\
% \authormark{2}Publications Department, Optica Publishing Group, 2010 Massachusetts Avenue NW, Washington, DC 20036, USA\\
% \authormark{3}Currently with the Department of Electronic Journals, Optica Publishing Group, 2010 Massachusetts Avenue NW, Washington, DC 20036, USA}

% \email{\authormark{*}takasumi@elec.keio.ac.jp} %% email address is required; see note below about the corresponding author designation
% \begin{abstract*} 

% \end{abstract*}

%%%%%%%%%%%%%%%%%%%%%%%%%%  Common body part %%%%%%%%%%%%%%%%%%%%%%%%%%
\section{Introduction}\label{sec1}
 Passively mode-locked pulsed lasers have emerged as essential tools in a variety of applications, including ultrafast laser spectroscopy and coherent control \cite{Zewail1988-ov,Itakura2003-tq}, femtosecond laser processing \cite{Korte2003-ij,Tanabe2005-hx}, attosecond physics \cite{Krausz2009-an,Gaumnitz2017-ve}, time and frequency standards \cite{Fortier2019-da}, and telecommunications \cite{Hasegawa1995-zb}. Shortly after the invention of the laser, research into short-pulse generation was conducted \cite{DeMaria1966-yc}, and mode-locking using saturable absorption (SA) was developed. Early laser systems employed laser dyes \cite{Ippen1972-tu} and semiconductor saturable-absorber mirrors (SESAMs)~\cite{Keller1990-ng}, both of which rely on natural SA. For a long time, Kerr-lens mode-locking \cite{Spence1991-vy}, a technique utilizing the self-focusing effect to create artificial SA, has been used for mode-locking in solid-state lasers. In recent years, the emergence of fibre lasers, which capitalize on both natural and artificial SA effects achieved through carbon nanotubes (CNTs) \cite{Set2004-jm}, graphene \cite{Bao2009-mw}, nonlinear polarization rotation \cite{Fermann1993-ol}, and nonlinear loop mirrors \cite{Duling1991-eq}, has made ultrashort pulse lasers more accessible.

 In recent times, the demand for high-repetition pulsed lasers has surged~\cite{Tanabe2019-uf,Kippenberg2018-fi,Ning2019-pj,Pfeifer2022-dv} as they enable more effective data acquisition and faster material processing speed. Consequently, researchers have developed harmonic mode-locking techniques in fibre lasers \cite{Collings1998-if}, short-cavity solid-state lasers \cite{Kimura2019-of}, mode locked integrated external-cavity surface-emitting lasers \cite{Keller2006-iw}, and $\mathrm{Er^{3+}}$ doped fibre-ferrule cavities \cite{Martinez2011-ed}. Among these, nanophotonic devices fabricated from erbium-ion ($\mathrm{Er^{3+}}$) doped glass \cite{Collot1993-pw,Armani2003-ee} stand out as potential cost-effective solutions, given the maturity and robustness of the material. These devices are particularly promising if a small mode-locked cavity can be fabricated. Therefore, a mode-locked laser using a whispering-gallery-mode (WGM) microresonator \cite{Prugger_Suzuki2021-wi}, could provide a cost-effective on-chip solution for demonstrating ultra-high repetition rate optical pulses. However, the realization of SA in such a system, without sacrificing essential parameters such as the cavity quality factor ($Q$), poses a challenge if we choose to utilize CNTs or graphene. Hence, it is essential to develop an entirely new approach for achieving artificial SA and enabling passive mode-locking, especially if we aim to realize femtosecond pulse laser sources in nanophotonic devices.

 At the same time, systems consisting of coupled resonators with gain and loss have been garnering interest and are being investigated for their ability to break parity-time (PT) symmetry in structures such as waveguides \cite{Ruter2010-eo}, photonic crystals \cite{Takata2021-xc}, and microrings\cite{Hodaei2014-zu,Peng2014-ry,Komagata2021-my}. It has been proven that a coupled cavity resonator can achieve low-threshold lasing when operated beyond exceptional points (EPs), where eigenfrequencies coalesce \cite{Peng2014-pc}. This is due to a significant reduction in system loss. Notably, the system loss in such configurations exhibits considerable modulation near an EP.

 In this work, we theoretically propose and numerically demonstrate passive mode-locking in a coupled-cavity system with gain and loss. This system uniquely leverages properties that emerge near the EPs. By exploiting the extensive system loss modulation that occurs close to the EP, we achieve efficient artificial SA, leading to ultrahigh-repetition mode-locking in a small cavity system. We also anticipate that the required quality factor ($Q$) of the system will be substantially relaxed thanks to the separation of gain and loss in two distinct cavities. Moreover, we introduce a new parameter: the coupling rate between the two microcavities. This additional control point should facilitate easier experimental demonstration.

\section{Coupled cavities for mode-locking}\label{sec2}
\subsection{Concept and operational mechanism}\label{subsec2-1}
\begin{figure}[ht]
\centering
\includegraphics[scale=0.9]{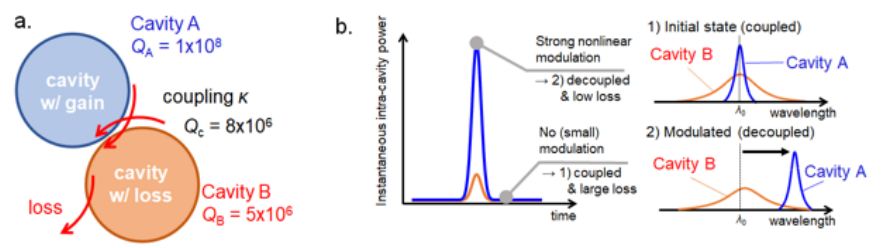}
\caption{Model and principle. \textbf{a.} Schematic representation of our model. Cavity A is a cavity with gain; Cavity B is a lossy cavity. Light is generated in Cavity A and undergoes loss in Cavity B. \textbf{b.}  Explanation of the principle of artificial SA in a coupled cavity system. The left graph shows the temporal waveform in Cavity A (blue line) and Cavity B (red line). The pulse intensity in Cavity A is consistently higher due to its gain properties. At the pulse's peak, Cavity A exhibits strong nonlinearity, leading to modulation in the refractive index. In contrast, the tail (or background) remains linear. The right graphs show the resonant spectra for both cavities at the peak and tail of the pulse. Initially, at the tail of the pulse, Cavities A and B couple with each other since they share the same resonant wavelengths. However, as the peak of the pulse in Cavity A demonstrates significant optical nonlinearity, it causes a shift in resonant wavelengths between the two cavities, resulting in their decoupling. Consequently, the light in Cavity A experiences reduced loss.}\label{fig-schematic}
\end{figure}

Our model system, as depicted in Fig.~\ref{fig-schematic}(a), is a pair of coupled microresonators . The first resonator, cavity A, provides gain, while the second one, cavity B, is a passive microresonator with the same diameter as cavity A but with a higher loss (low $Q$). The difference between the $Q$ values of the two resonators should be noted because the distinction is crucial as regards the mode-locking mechanism explored in this study.

With gain provided by cavity A, single and multi-mode lasing can be achieved when a pump is applied. However, continuous wave (CW) lasing is only obtained in the absence of an SA effect. Figures~\ref{fig-schematic}(a) and 1b show how our system can exhibit an artificial SA that contributes to pulsing. Given the coupled nature of cavities A and B, and the lossy attribute of cavity B, the light from cavity A couples into cavity B, leading to its elimination due to the low-$Q$; an observation suggesting that we can reduce system loss if we can manage the decoupling of the two cavities.

The influence of the Kerr effect, characterized by an intensity-dependent, instantaneous refractive index change, such as,

\begin{equation}\label{kerr}
    n = n_0 +n_2 I ,
\end{equation}
where $n_2$ is the nonlinear refractive index, and $I$ the intensity, is evident when examining a pulse circulating within the cavity. With gain in cavity A and $Q_\mathrm{A} > Q_\mathrm{B}$, the pulse peak shows pronounced nonlinear refractive index modulation in cavity A, causing disparate resonant frequency modulations in the two cavities, as explained in Fig.~\ref{fig-schematic}(b). This results in an instantaneous decoupling of the cavities at the pulse peak; this is an intuitive explanation despite the challenge of defining an instantaneous resonant frequency within a round trip. Rigorously, this decoupling can also be viewed as a phase velocity mismatch caused by the Kerr effect. With initially identical resonant frequencies, cavities A and B experience modulation and decoupling only at the pulse peak. As a consequence, the light within cavity A undergoes decreased loss at the pulse peak, thereby replicating the operation of SA behaviour.

The primary challenge is to secure a substantial degree of coupling/decoupling modulation under conditions of minor refractive index modulation, given that $n_2$ is typically small. To surmount this challenge, we propose exploiting the potent modulation commonly observed near an EP in a coupled microresonator system.

\subsection{Numerical model}\label{subsec2-2}
To investigate the new mode-locking technique that forgoes the use of any natural SAs, we have constructed a model utilizing the approach described below with nonlinear Schr\"{o}dinger equations with gain~\cite{Prugger_Suzuki2021-wi} and mode coupling terms~\cite{Fujii2017-yr}, such as,

\begin{equation}\label{LLE-A}
\begin{split}
    T_\mathrm{r} \frac{\partial A\left ( t, T\right )}{\partial T} = iL \left( - \frac{\beta _2}{2}\cdot \frac{\partial^2}{\partial t^2} + \gamma_{\mathrm{nl}} \left| A \left ( t, T\right ) \right|^2 \right ) A\left( t,T\right) + \frac{{g_T}_\mathrm{r}\left( T\right) -l_\mathrm{A}}{2} A\left ( t, T\right ) + i\frac{\kappa}{2}B\left( t,T \right) ,
\end{split}
\end{equation}

\begin{equation}\label{LLE-B}
\begin{split}
    T_\mathrm{r} \frac{\partial B\left ( t, T\right )}{\partial T} = iL \left( - \frac{\beta _2}{2}\cdot \frac{\partial^2}{\partial t^2} + \gamma_{\mathrm{nl}} \left| B \left ( t, T\right ) \right|^2 \right ) B\left( t,T\right) - \frac{l_\mathrm{B}}{2} A\left ( t, T\right ) + i\frac{\kappa}{2}A\left( t,T \right) ,
\end{split}
\end{equation}
where $A(t,T)$ and $B(t,T)$ are the complex electrical field amplitudes (in a slowly-varying envelope approximation) in cavities A and B. Here, $t$ and $T$ are the fast time and the slow time, respectively. Equations (\ref{LLE-A}) and (\ref{LLE-B}) are interconnected through the last term, with $\kappa = \omega_0 / Q_\mathrm{c}$ representing the coupling rate, where $Q_\mathrm{c}$ is the coupling $Q$. $\gamma_\mathrm{nl} ( = n_2 \omega_0 (cA_\mathrm{eff} )^{-1})$ is the nonlinear coefficient. ${g_T}_\mathrm{r}(T)$, and $l_\mathrm{A,B}$ are the gain per roundtrip in cavity A, and losses per roundtrip in cavities A or B, as below \cite{Prugger_Suzuki2021-wi}, 

\begin{equation}\label{gain}
    {g_T}_\mathrm{r} = g_0 \left ( \frac{1}{1 + \frac{\overline{ \left | A(t,T) \right |}^2}{P^g _\mathrm{sat}}} \right ) \left ( 1 + \frac{1}{\Delta f_g ^2} \frac{\partial^2}{\partial t^2} \right ) ,
\end{equation}

\begin{equation}\label{loss}
    l_\mathrm{A,B} = T_\mathrm{r} \gamma_\mathrm{A,B} = T_\mathrm{r}\cdot \frac{\omega_0}{Q_\mathrm{A,B}} ,
\end{equation}
where the parameters are described in Tab.~\ref{table1}.
When $A$ is nonzero, the last term on the right-hand side of Eq.~(\ref{LLE-A}) can be rewritten as follows, 

\begin{equation}\label{complex}
    i \frac{\kappa}{2}B = \left [ \mathrm{Real} \left ( i\frac{\kappa}{2}\frac{B}{A} \right ) + i\cdot \mathrm{Imag} \left ( i\frac{\kappa}{2}\frac{B}{A} \right ) \right ] A ,
\end{equation}
where the real part (the first term) corresponds to gain and loss, whereas the imaginary part (the second term) pertains to the phase modulation, both arising due to the coupling. Given that $A$ and $B$ are functions of time $t$, Eq.~(\ref{complex}) shows that the effective gain and loss emerging as a result of the coupling constitute a time-dependent function and may act as an artificial SA.

Equations (\ref{LLE-A}) and (\ref{LLE-B}) are solved using the split-step Fourier method, which is a technique widely adopted for calculations involving microresonator frequency combs~\cite{Fujii2017-yr}. As laid out in Tab.~\ref{table1}, we adopt a silica toroid microresonator with a diameter of $D = 300$~\textmu m for our model. We only doped cavity A with $\mathrm{Er^{3+}}$ ions to produce gain, setting the $Q_\mathrm{A}$ and $Q_\mathrm{B}$ values at $1 \times 10^8$ and $5 \times 10^6$, respectively. Cavities A and B are coupled with a coupling Q of $Q_\mathrm{c} = 8 \times 10^6$. Parameters for dispersion and mode area are obtained from the given structure, with the minor diameter of the toroid specified as 30~\textmu m. We note that both cavity exhibit the identical anomalous dispersion value ($\beta_2 = -12.24~\mathrm{ps^2/km}$).
%This consequently reveals the dispersion within our resonator system to be anomalous in nature.

\section{Results and Discussion}\label{sec3}
\subsection{Demonstration of the mode-locking}\label{sec3-1}
\begin{figure}[ht]
\centering
\includegraphics[scale=0.85]{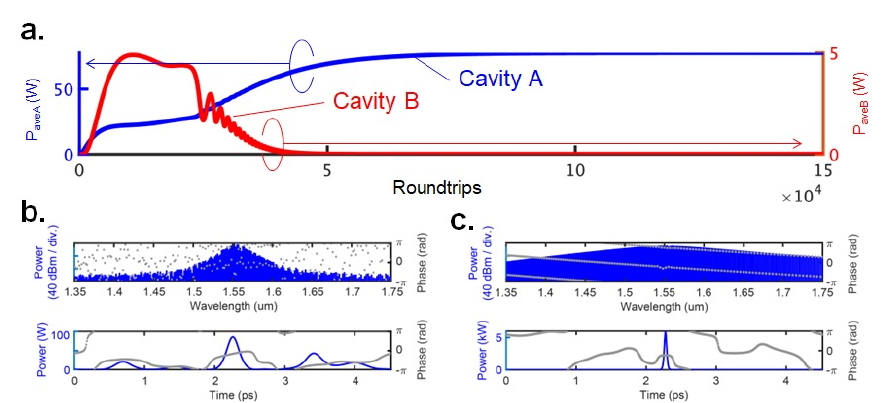}
\caption{Calculation results obtained from coupled nonlinear Schrödinger equations. \textbf{a.} The average optical power in Cavity A (blue) and Cavity B (red) as a function of the roundtrips. \textbf{b.} Spectral intensity and phase, alongside temporal intensity and phase, within a roundtrip at roundtrip 10,000. \textbf{c.} At roundtrip 30,000.}\label{fig-regime}
\end{figure}

\begin{figure}[ht]
\centering
\includegraphics{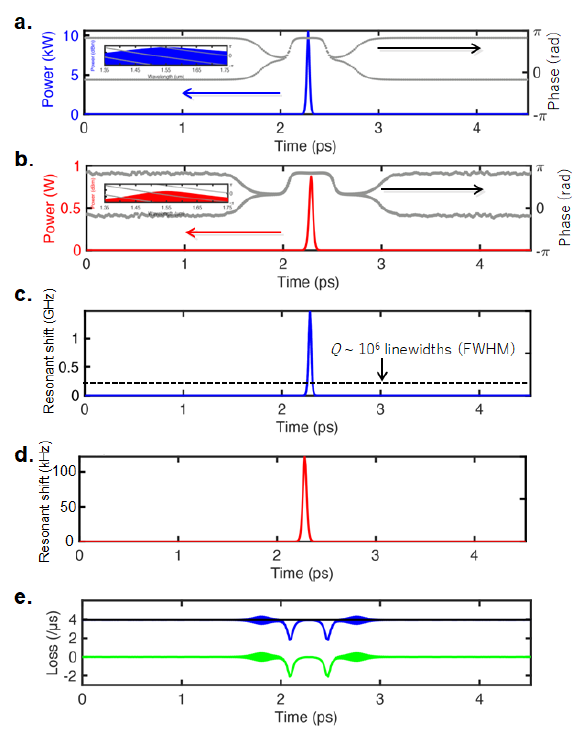}
\caption{\textbf{a.} Mode-locked temporal waveform and spectrum (inset) in Cavity A at roundtrip 70,000. The blue line represents intensity, while the grey lines show the phase. \textbf{b.} As \textbf{a.}, but for Cavity B. \textbf{c.} Calculated resonant shift for Cavity A resulting from Kerr nonlinear refractive index modulation. A maximum frequency shift of $\sim 1.5$~GHz is observed at the pulse peak. The black dotted line indicates the frequency linewidth (FWHM) for a cavity with a $Q$ of $10^6$. \textbf{d.}  As \textbf{c.} but for Cavity B. \textbf{e.} The nonlinear loss and gain profile in Cavity A arising from the system's coupling and decoupling within the roundtrip. The black solid line depicts the net gain in Cavity A (i.e., $g_0$ minus the loss $Q_\mathrm{A}$). The green line shows the extra loss in Cavity A due to its coupling with Cavity B. The blue curve portrays the overall net gain/loss profile in Cavity A. The green and blue curves are applicable only when the intensity within cavity A is nonzero.}\label{fig-LLEresult}
\end{figure}

The results of our computation are shown in Fig.~\ref{fig-regime}(a), where we chart the average power within each cavity ($P_\mathrm{aveA}$ and $P_\mathrm{aveB}$) as a function of the number of roundtrips after starting the cavity pump. As multimode lasing begins, the longitudinal modes remain unlocked, and the temporal waveform undergoes a random modulation (Fig.~\ref{fig-regime}(b). Following several thousand round trips, we begin to observe the formation of a pulse, albeit an unstable one (Figs.~\ref{fig-regime}(c)). It takes approximately $7 \times 10^4$ round trips for the cavity to reach a state of complete stability. The temporal waveforms for cavities A and B are shown in Figs.~\ref{fig-LLEresult}(a) and \ref{fig-LLEresult}(b), respectively. At this juncture, we witness clear mode-locking behaviour. The pulse shape has a $\mathrm{sech^2}$ profile with a full-width at half maximum (FWHM) of 30~fs. The phase of the pulse reveals it to be Fourier transform-limited, providing direct evidence of mode-locking, with a corresponding spectral width of 84~nm (FWHM).

Here, we confirm that the Kerr effect instigates instantaneous decoupling, and thus serves as an artificial SA. Figures~\ref{fig-LLEresult}(c) and \ref{fig-LLEresult}(d) show the associated resonant wavelength shifts for the waveforms shown in Figs.~\ref{fig-LLEresult}(a) and \ref{fig-LLEresult}(b), calculated via instantaneous nonlinear refractive index modulation as given in Eq.~(\ref{kerr}). The outcome indicates that the resonance of cavity A effectively undergoes a frequency shift of up to 1.5~GHz. This shift surpasses the linewidth of cavity B (with $Q_\mathrm{B} = 5 \times 10^6$). Given that the shift in cavity B is negligible (Fig.~\ref{fig-LLEresult}(d)), we can instantaneously decouple cavity A from cavity B only at the pulse peak. This process results in minimal loss and allows the presence of the auxiliary cavity B to function as an artificial SA.

Figure~\ref{fig-LLEresult}(e) offers greater clarity by depicting the effective gain and loss experienced by cavity A due to its coupling with cavity B within a roundtrip. The loss is clearly negligible at the pulse peak and subsequently increases at the pulse tail. This finding reinforces our previous conclusion that the coupled cavity system indeed realizes an artificial SA.
However, a lingering question persists: why does the SA occur so effectively with this parameter set? Generally, the $n_2$ value is small, implying that the modulation would only occur to a limited extent. To answer this question, it becomes crucial to exploit the impact of the EP within this system.

\subsection{The impact of the exceptional point}\label{sec3-2}
Upon a closer examination of the temporal phase in Fig.~\ref{fig-LLEresult}(b), a unique structure is observable where the phase bifurcates before consolidating into one as the intensity ascends. In an attempt to comprehend this behaviour, we undertake a steady-state analysis of our coupled cavity system.

To simplify the picture, we consider two cavities; one with low loss (cavity A, with $Q_\mathrm{A} = 10^8$) and the other with high loss (cavity B, with $Q_\mathrm{B} = 5 \times 10^6$). A coupled-mode analysis is then performed based on the following expressions~\cite{Peng2014-pc}:

\begin{equation}\label{CMT-A}
    \frac{dA}{dt} = i\omega_\mathrm{A}A - \frac{\gamma_\mathrm{A}}{2} A + i\frac{\kappa}{2}B ,
\end{equation}

\begin{equation}\label{CMT-B}
    \frac{dB}{dt} = i\omega_\mathrm{B}B - \frac{\gamma_\mathrm{B}}{2} B + i\frac{\kappa}{2}A .
\end{equation}

At a steady state, the complex eigenfrequencies are given as

\begin{equation}\label{eigen}
    \omega_\pm = \frac{\omega_\mathrm{A} + \omega_\mathrm{B}}{2} + i\frac{\chi}{2} \pm \beta ,
\end{equation}
where, $\chi$ is the average decay rate given by $\chi = \frac{1}{2}\left ( \gamma_\mathrm{A} + \gamma_\mathrm{B} \right )$. Here, we introduce $\beta (= \frac{1}{2}\sqrt{\kappa^2 - \Gamma^2}) $ to represent the effect of the coupling rate $\kappa$ and of the difference in the decay rates of the two cavities, given by $\Gamma = \frac{1}{2}\left ( -\gamma_\mathrm{A} + \gamma_\mathrm{B} \right )$. It is worth noting that the real part of Eq.~(\ref{eigen}) corresponds to the resonant frequencies, whereas the imaginary part signifies the system loss (i.e., $Q$).

\begin{figure}[ht]
\centering
\includegraphics[scale=0.65]{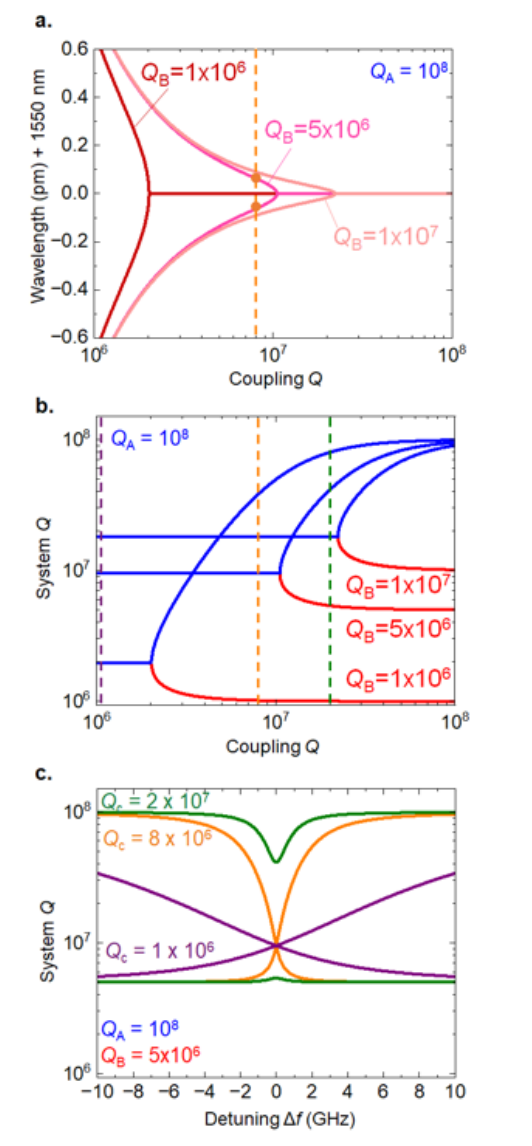}
\caption{\textbf{a.} The real part of the eigenfrequency (resonant frequency) depicted as a wavelength shift from the baseline resonance at 1550~nm, for various $Q_\mathrm{B}$ values. $Q_\mathrm{A}$ is consistently set at $10^8$. The orange line represents the condition where $Q_\mathrm{c} = 8 \times 10^6$, the parameter set utilized for the computations in Figs. \textbf{2} and \textbf{3}. \textbf{b.} Imaginary part of the eigenfrequency (representing gain and loss) displayed as system $Q$, across different $Q_\mathrm{B}$ values. $Q_\mathrm{A}$ remains constant at $10^8$. \textbf{c.} System $Q$ plotted against the detuning between Cavity A and Cavity B for a range of $Q_\mathrm{c}$ values. $Q_\mathrm{A}$ and $Q_\mathrm{B}$ are set at $10^8$ and $5 \times 10^6$, respectively. The condition $Q_\mathrm{c} = 8 \times 10^6$ is consistent with the parameters employed in the analyses of Figs.\textbf{2}  and \textbf{3}.}\label{fig-eigenfreq}
\end{figure}

Figures~\ref{fig-eigenfreq}(a) and \ref{fig-eigenfreq}(b) present the imaginary and real components of the eigenfrequencies stipulated in Eq.~(\ref{eigen}), plotted as functions of $Q_\mathrm{c}$, which is the coupling $Q$ between the two cavities. The plots vary with different $Q_\mathrm{B}$ values. Figure~\ref{fig-eigenfreq}(a) shows that the eigenfrequency (expressed in wavelength) moves beyond the EP as $Q_\mathrm{c}$ increases. Here we consider the unique phase bifurcations observed in Fig.~\ref{fig-LLEresult}(a), which we calculated using a $Q_\mathrm{B}$ value of $5 \times 10^6$. The system is linear at the tail of the pulse and resides below the EP because $Q_\mathrm{c}$ is at $8 \times 10^6$, as indicated by the orange dotted line in Fig.~\ref{fig-eigenfreq}(a). Owing to the bifurcation of the real part of the eigenfrequency, a phase split becomes observable. As the pulse intensity increases, $Q_\mathrm{c}$, quite intuitively, also effectively grows thanks to the increased decoupling, leading to a collision of the eigenfrequencies. Consequently, their phases merge into a single phase.

In Fig.~\ref{fig-eigenfreq}(b), we witness substantial modulation of the system $Q$ in proximity to the EP in all instances, a feature that substantially contributes to the efficient SA behaviour. This outcome implies that if we can modulate $Q_\mathrm{c}$, we might be able to adjust the system $Q$ significantly and consequently realize an effective SA. 
% However, in the experiment, we need to fix $Q_\mathrm{c}$ and this means we are only able  to modulate the refractive index of cavity A using the Kerr effect. To this end, we fixed $Q_\mathrm{B}$ at $5 \times 10^6$ and investigated the impact of the detuning.

Figure~\ref{fig-eigenfreq}(c) shows the system $Q$ as a function of the frequency detuning $\Delta f$ between cavities A and B (i.e., $\Delta f = f_\mathrm{A}- f_\mathrm{B}$) while maintaining $Q_\mathrm{B}$ at $5 \times 10^6$. It shows results for various $Q_\mathrm{c}$ values. If $Q_\mathrm{c}$ is set at $2 \times 10^7$, the system is beyond the EP, which signifies that the two modes diverge into high and low $Q$s from the outset (as depicted by the green dotted line in Fig.~\ref{fig-eigenfreq}(b)). Consequently, the system $Q$s have two values from the beginning, and the modulation remains shallow, even when we alter the refractive index of the cavity. Conversely, if $Q_\mathrm{c}$ is too small at $Q_\mathrm{c} = 1 \times 10^6$, the starting point is significantly below the EP, as portrayed by the purple dotted line in Fig.~\ref{fig-eigenfreq}(b). To obtain efficient modulation, we must initialize close to the EP, specifically at $Q_\mathrm{c} = 8 \times 10^6$, which is just below the EP. As a result, with minor modulation of the detuning between the cavities, we can obtain significant modulation of the system $Q$.

While our operation is set just below the EP, it is equally possible to initialize the system just above the EP. However, it is critical to remain close to the EP to ensure effective modulation of the system $Q$. The resonant shift that we achieve is approximately 1.5~GHz (as shown in Fig.~\ref{fig-LLEresult}(c)), leading to a modulation of the system $Q$ from $\sim 1 \times 10^7$ to $\sim 4 \times 10^8$, according to Fig.~\ref{fig-eigenfreq}(c).

\subsection{Mode-locking range} \label{sec3-3}
\begin{figure}[ht]
\centering
\includegraphics[scale=0.7]{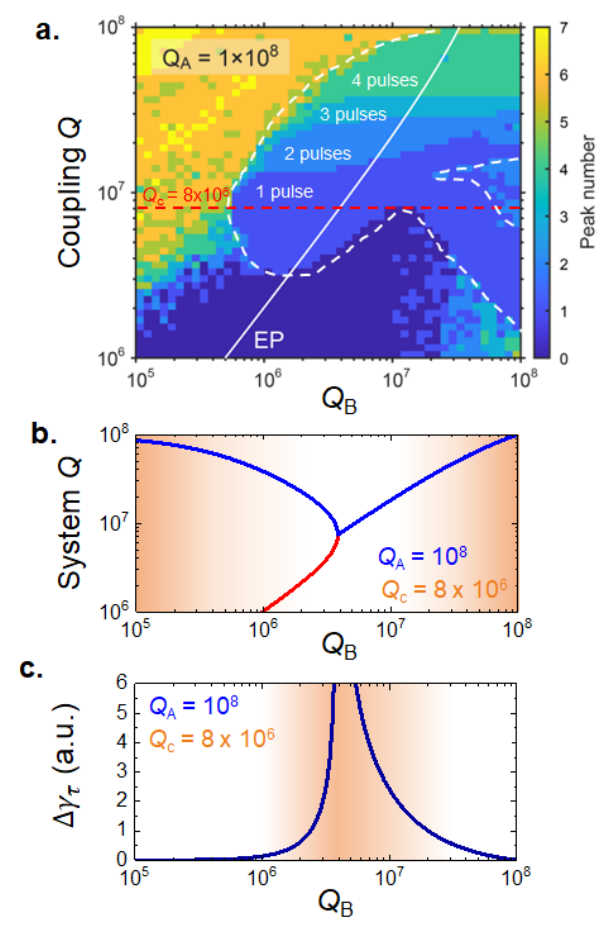}
\caption{\textbf{a.} Colour map illustrating the count of peaks in the temporal waveform per roundtrip (indicating the number of pulses per roundtrip) after 70,000 roundtrips as a function of $Q_\mathrm{B}$ and $Q_\mathrm{c}$. The region enclosed by a white dotted line represents the stable mode-locked regime. The white solid line signifies the position of the EP. \textbf{b.} System $Q$, as calculated from Eq. (9), plotted as a function of $Q_\mathrm{B}$ when $Q_\mathrm{A}$ and $Q_\mathrm{c}$ are $10^8$ and $8 \times 10^6$, respectively. The system $Q$ decreases as $Q_\mathrm{B}$ reduces, but it recovers due to the presence of the EP. A high $Q$ is essential to obtain sufficient net gain and optical nonlinearity in cavity A, which are required for lasing and mode-locking. \textbf{c.} $\Delta\gamma_\tau$ plotted as a function of $Q_\mathrm{B}$ when $Q_\mathrm{A}$ and $Q_\mathrm{c}$ are $10^8$ and $8 \times 10^6$, respectively. $\Delta\gamma_\tau$ reaches its maximum value of $\sim 34.2$ in an EP condition. The definition of $\Delta\gamma_\tau$ is provided in the main text. A higher value corresponds to substantial modulation of the system loss rate given a specific refractive index modulation. This property is necessary for achieving efficient artificial SA.}\label{fig-map}
\end{figure}

Lastly, we undertake a search for the optimal parameters. Figure~\ref{fig-map}(a) presents our findings, visually encapsulated by using a colour map depicting the number of peaks in the waveform generated after 70,000 roundtrips. Furthermore, we evaluated the stability of the resulting waveform in an attempt to distinguish between stable mode-locking, breather dynamics, and chaotic behaviour. The region of stability is denoted by a white-dotted contour. The white solid line indicates the condition of the EP.

To understand the result, we begin by pointing out that two conditions are needed to achieve mode-locking. The first condition is a high system $Q$ because a low loss is needed to achieve lasing in the first place. It also helps facilitate efficient optical nonlinear effects. The second condition is to achieve efficient SA behaviour, which is essential for mode-locking. We employ Figs.~\ref{fig-map}(b) and \ref{fig-map}(c) to investigate these two properties.

Figure~\ref{fig-map}(b) is the system $Q$ as a function of $Q_\mathrm{B}$ when $Q_\mathrm{A} = 10^8$ and $Q_\mathrm{c} = 8 \times 10^6$. The system $Q$ is high when $Q_\mathrm{B}$ is high since most of the light exhibits loss through cavity B. But the system $Q$ recovers even when $Q_\mathrm{B}$ exhibits a reduction in its value due to the presence of the EP. Figure~\ref{fig-map}(c) presents the maximum slope value of the system loss rate $\Delta \gamma_\tau$ across varying $Q_\mathrm{B}$ values. The system $Q$ ($Q_\mathrm{sys}$) as a function of the cavity detuning $\Delta f$ was previously outlined in Fig.~\ref{fig-eigenfreq}(c). The system loss rate, $\gamma_\tau$ is represented by $\gamma_\tau = \omega_0 / Q_\mathrm{sys}$, which also varies as a function of detuning $\Delta f$. Thus, we define $\Delta \gamma_\tau$ as,

\begin{equation}\label{gamma}
    \Delta \gamma_\tau = max \left | \frac{d\gamma_\tau(\Delta f)}{d(\Delta f)}\right |.
\end{equation}

Here, $max$ indicates that we extract the peak value of the function. Computing $\Delta\gamma_\tau$ for different $Q_\mathrm{B}$ values provides the plot illustrated in Fig.~\ref{fig-map}(c). Intuitively, a larger value of $\Delta\gamma_\tau$ indicates that even small detuning, arising from the Kerr effect, can lead to significant modulation in the system's loss rate. In essence, a substantial $\Delta\gamma_\tau$ is indicative of an efficient artificial SA.

To gain a comprehensive understanding of the results presented in Fig.~\ref{fig-map}(a), let us consider a fixed $Q_\mathrm{c}$ at $8 \times 10^6$. Initiating our discussion at high $Q_\mathrm{B}$ values, where $Q_\mathrm{B}$ is $\sim 10^8$, the system $Q$ is at its zenith. However, as suggested by Fig.~\ref{fig-map}(c), the potency of the artificial SA effect is minimal, rendering the system unstable. As $Q_\mathrm{B}$ falls to roughly $5 \times 10^7$, the increase in $\Delta\gamma_\tau$ facilitates mode-locking in the system. The further descent of $Q_\mathrm{B}$ to around $\sim 10^7$ situates the system on the fringes of stable mode-locking, as depicted in Fig.~\ref{fig-map}(a). Even as $\Delta\gamma_\tau$ increases, the system $Q$ diminishes substantially. It is important to note that $Q$ values degenerate beyond the EP (to the right of the solid white line in Fig.~\ref{fig-map}(a)). This infers that the intensities in cavities A and B are nearly identical, implying that the nonlinear refractive index modulations are virtually equivalent. Consequently, efficient detuning between cavities A and B becomes unattainable, and therefore the recovered $Q$ hinders the stable mode-locked operation (i.e., breathing behaviour). Yet, as $Q_\mathrm{B}$ nears the EP, $\Delta\gamma_\tau$ experiences a notable surge, peaking at approximately 34.2 at the EP. Owing to the pinnacle slope that occurs at $\Delta f = 0$ , significant artificial SA arises close to EP, facilitating mode-locking even with diminished system $Q$ values.

A region warranting particular attention lies to the left of the EP condition ($Q_\mathrm{B} < 4 \times 10^6$). Here, while $\Delta\gamma_\tau$ undergoes a decline, the system Q exhibits a considerable resurgence, facilitating mode-locking courtesy of the pronounced Kerr effect. In contrast to the region to the right of the EP, the system $Q$ between modes A and B shows a marked disparity that augments efficient system decoupling. This stability in mode-locking persists until $Q_\mathrm{B}$ falls to $6 \times 10^5$. Beyond this, the diminished $\Delta\gamma_\tau$ restricts the attainment of mode-locking (for $Q_\mathrm{B} < 6 \times 10^5$).

The insights gleaned from Figs.~\ref{fig-map}(b) and \ref{fig-map}(c) shed light on the restoration of mode-locking as $Q_\mathrm{B}$ decreases, a process that can be attributed to the unique behaviour of the EP in a coupled cavity system. This demonstrates the role of the EP as an enabler of mode-locking even under low-$Q_\mathrm{B}$ conditions, revealing novel avenues for the manipulation and control of lasing dynamics. Significantly, the optimal region for mode-locking is relatively broad, thus offering promising prospects for experimental implementation.

\section{Conclusion}\label{sec4}
In conclusion, our study provides a comprehensive examination of a novel mode-locking strategy that forgoes the utilization of natural SAs, instead leveraging the Kerr effect and a coupled microresonator system in close proximity to an EP. The results reveal the occurrence of mode-locking after approximately $7 \times 10^4$ round trips. The formed pulse adopts a $\mathrm{sech^2}$ shape, with its Fourier transform-limited phase. The resonant shift associated with the light intensity modulation underscores the role of instantaneous decoupling as an artificial SA, contributing to efficient mode-locking. Our exploration of the parameter space further clarifies the conditions under which mode-locking is achieved. Notably, mode-locking consistently occurs close to the EP, where substantial modulation of the system $Q$ is possible.

Our research opens new avenues for exploring coupled microresonator systems as platforms for efficient mode-locking, moving beyond the traditional use of saturable absorbers. The broad sweet spot for achieving mode-locking, identified through our investigations, offers exciting possibilities for the experimental validation and application of this technique.

\newpage
% -------------------------------- Figures --------------------------------
\section*{Figures}
\subsection*{Figure 1}

\newpage
\subsection*{Figure 2}

\newpage
\subsection*{Figure 3}

\newpage
\subsection*{Figure 4}

\newpage
\subsection*{Figure 5}

\clearpage
\newpage
\textbf{Table}
% -------------------------------- Table. 1. --------------------------------
\begin{table}[ht]
\caption{Parameters employed in the calculations unless otherwise specified in the text\label{table1}}
    \centering
    \begin{tabular}{l c c c c}
    \hline
    Parameter                   & Variable              & Value of A                & Unit\\
    \hline
    Intrinsic $Q$ of Cavity A   & $Q_{\mathrm{A}}$      & $1\times 10^8$            & $-$\\
    Intrinsic $Q$ of Cavity B   & $Q_{\mathrm{B}}$      & $5\times 10^6$            & $-$\\
    Coupling Q                  & $Q_{\mathrm{c}}$      & $8\times 10^6$            & $-$\\
    Cavity diameter             & $D (= L/\pi)$         & 300                       & \textmu m\\
    Resonant wavelength         & $\lambda_0 (= 2\pi c/\omega_0)$ & 1.55            & \textmu m\\
    Roundtrip time              & $T_\mathrm{r}$        & 4.52                      & ps\\
    Refractive index            & $n_0$                 & 1.44                      & $-$\\
    Nonlinear refractive index  & $n_2$                 & $2.2 \times 10^{-20}$     & $\mathrm{m^2 / W}$\\
    Effective mode area         & $A_\mathrm{eff}$      & $21.28$                   & \textmu$\mathrm{m^2}$\\
    Second order dispersion     & $\beta_2$             & $-12.24$   & $\mathrm{ps^2/km}$\\
    Saturated gain              & $g_0$                 & $5 \times 10^{-2}$        & $/\mathrm{roundtrip}$\\
    Saturation power            & $P^\mathrm{g}_\mathrm{sat}$ & 144.8               & mW\\
    Gain bandwidth              & $\Delta f_\mathrm{g}$ & 2.50                    & THz\\
    \hline
    \end{tabular}
\end{table}

% -------------------------------- Acknowledgements --------------------------------
\textbf{Acknowledgments}
This work was supported by JSPS KAKENHI (JP19H00873, JP22K14625), by MEXT Quantum Leap Flagship Program (MEXT Q-LEAP) JPMXS0118067246 and by JST SPRING (JPMJSP2123).

\textbf{Competing financial interests} 
The authors declare that they have no competing financial interests.

\textbf{Correspondence} and requests for material should be addressed to TT (email: takasumi@elec.keio.ac.jp).

%%%%%%%%%%%%%%%% For optica template
% \begin{backmatter}
% \bmsection{Funding}

% \bmsection{Acknowledgments}

% \bmsection{Disclosures}
% The authors declare no conflicts of interest.

% \bmsection{Data availability} Data underlying the results presented in this paper are not publicly available at this time but may be obtained from the authors upon reasonable request.

% \bmsection{Supplemental document}
% See Supplement 1 for supporting content. 

% \end{backmatter}
%%%%%%%%%%%%%%%
\newpage
%%%%%%%%%%%%%%%%%%%%%%% References %%%%%%%%%%%%%%%%%%%%%%%%%
%%%%%%%%%% If using BibTeX:
\bibliography{EP_ref}
\bibliographystyle{opticajnl}

\end{document}